\documentclass[twocolumn,showpacs,preprintnumbers,amsmath,amssymb, prl]{revtex4}

\usepackage{graphicx}
\usepackage{dcolumn}
\usepackage{bm}

\begin{document}

\preprint{APS/123-QED}

\title{Waveguiding and Plasmon Resonances in Two-Dimensional \\ Photonic Lattices of Gold and Silver Nanowires
}

\author{M. A. Schmidt}
\author{L. N. Prill Sempere}
\author{H. K. Tyagi}
\author{C. G. Poulton}
\author{P. St.J. Russell}
 \affiliation {Max-Planck Research Group (IOIP), University of Erlangen-Nuremberg, Guenther-Scharowsky-Str.1/Bau 24, 91058 Erlangen, Germany}

\date{\today}

\begin{abstract}
We report the fabrication of triangular lattices of parallel gold and silver nanowires of high optical quality, with diameters down to 500 nm and length-to-diameter ratios as high as 100,000. The nanowires are supported by a silica glass matrix and are disposed around a central solid glass core, i.e., a missing nanowire. These cm-long structures make it possible to trap light within an array of nanowires and characterize the plasmon resonances that form at specific optical frequencies. Such nanowire arrays have many potential applications, e.g., to imaging on the sub-wavelength scale.
\end{abstract}

\pacs{42.70.Qs, 42.81.Qb, 78.67.Bf, 73.20.Mf}
\maketitle

Surface-plasmon-polaritons form at metal-dielectric boundaries as a result of the intense photon-electron interactions that result when light encounters a metal. They are tightly bound to the interface because, unlike in normal dielectric waveguides, there is no middle ``guiding'' layer. This results in strong field enhancements, which in turn can be used to increase the sensitivity of optical sensors or the strength of nonlinear effects. On nano-scale metallic objects, surface-plasmon resonances (SPRs) form at frequencies where the oscillating fields are able to fit around the particle without discontinuities in phase.\\
Metallic nano-objects have been the subject of many experimental and theoretical studies in recent years. SPRs at visible wavelengths have been reported in chemically grown gold ``nanorods'' with diameters of $\sim \,20$ nm and aspect ratios of $ \sim \, 1:5 $ \cite{Pelton_2006}. Propagation distances of $ \sim \, 50 \mu $m have been achieved in nanoscopic coaxial cables made from a central carbon nanotube, an alumina filling and a Cr outer sheath \cite{Rybczynski_2007}. Planar nanowire arrays have also been produced by lithographic techniques \cite{Schider_2003}. Arrays of parallel nanowires have also been suggested as suitable structures for sub-wavelength imaging \cite{Elser_2006, Podolskiy_2005, Shvets_2007}. The optical and vibrational properties of silver particles have been studied experimentally using optical transmission spectroscopy and low-frequency Raman scattering \cite{Margueritat_2007}. \\
It has been suggested that SPRs have the potential to allow very-large-scale-integration of photonic devices at high packing densities \cite{Ozbay_2006, Maier_2006}. A major challenge, however, is how to create nano-scale metallic structures in which the impact of high optical absorption in the metal is reduced, while preserving the advantages of strong metal-light interactions. \\
Here we report a new class of waveguiding structure that goes some way to achieving this. It consists of a triangular array of parallel gold or silver nanowires, arranged around a central missing nanowire which acts as a waveguide and as a convenient means of probing the optical properties of the nanowire array. The arrays were formed by pumping molten gold and silver into the narrow hollow channels of ``holey'' photonic crystal fibers (PCFs) made from silica glass. The PCFs used had a central solid glass core surrounded by a triangular lattice of hollow channels. Three different hole diameters were studied (550 nm, 980 nm and 1520 nm) at a constant inter-hole spacing of $ \Lambda \approx \, 2.9 \mu $m, resulting in diameter-to-spacing ratios ($ d / \Lambda $) ranging from 0.18 to 0.50.

\begin{figure}[htb]
	\includegraphics*[width=6.9cm]{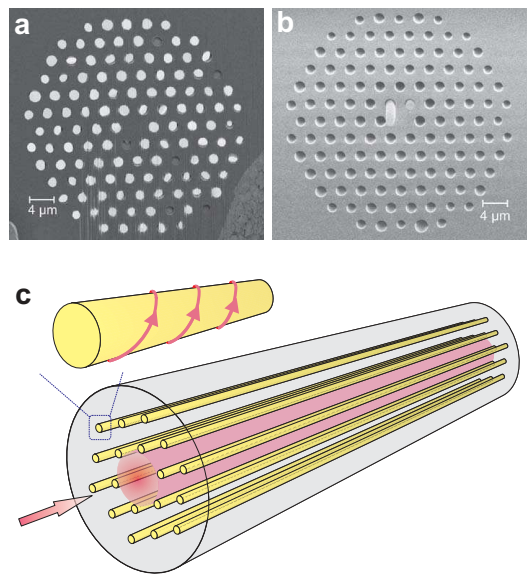}
	\caption{\label{fig1}(a) Scanning electron micrograph (SEM) of the end-face of a cleaved metal-filled PCF, polished by focused ion beam etching (hole diameter $ 1.52 \mu $m, hole spacing $2.9 \mu $m). The wires undergo ductile thinning during cleaving, breaking at random positions. Those that protrude from the end-face are polished and appear as bright disks in the SEM; the rest (6 in total) are dull in appearance. (b) SEM image of the unpolished end-face of a PCF with two metal nanowires located next to the solid core. (c) Schematic of a metal nanowire array; SPRs are created by surface plasmons spiralling around the nanowires.}
\end{figure}

\begin{figure}[htb]
	\includegraphics*[width=7cm]{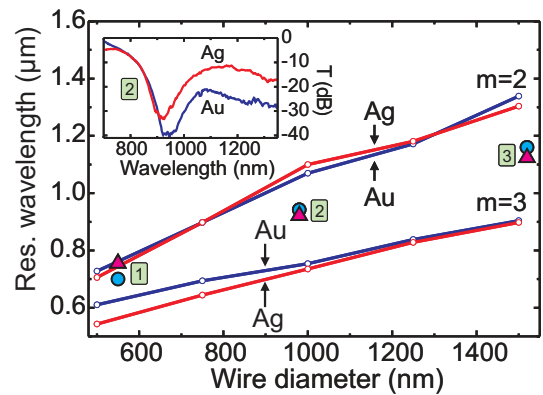}
	\caption{\label{fig2} SPR wavelength as a function of nanowire diameter. The white dots on the solid lines are the numerically simulated positions of the $m = 2$ and $m = 3$ SPRs, and the isolated points (triangles: silver; circles: gold) are experimental measurements. The points labeled 1 and 2 are for PCFs completely filled with metal, whereas point 3 is for a PCF in which only three holes were filled (two in the first ring and one in the second). The inset shows the typical transmission (T) spectra of PCFs filled with gold (blue curve) and silver (red curve). The wire diameter was 980 nm in each case, and the transmission was referenced to that of an unfilled PCF.}
\end{figure}

The hole-filling procedure involves melting the metal in a high-temperature ( $\approx \, 1100 ^\circ $C) pressure cell and pumping it into the hollow PCF channels at pressures of up to 60 bar. Penetration lengths up to 40 mm were routinely achieved. The PCF drawing process produces glass walls of extreme smoothness (roughness due to frozen-in capillary waves being a small fraction of a nanometer deep \cite{Roberts_2005}), resulting in nanowires of excellent optical quality. Another advantage of this approach, compared to other deposition or infiltration methods using organic precursors \cite{Finlayson_2007}, is that the wires are inherently free of impurities. \\
Figures \ref{fig1} (a) and (b) show representative SEMs of the end-faces of gold-filled PCFs. Because Au and Ag melt at temperatures well below the softening point of the glass, the initial PCF geometry is well preserved. The smallest nanowire diameter realized in the experiments was 550 nm. Single and multi-wire structures were successfully fabricated by fusing shut all but selected holes at the PCF end-face before pumping in the liquid metal (Fig. \ref{fig1}(b)).\\
Measurements of electrical conductivity of fibers with a single nanowire confirmed that the Au nanowires were continuous over the entire length of the filled section. The Ag wires, in contrast, showed cracks a few tens of $\mu $m wide, spaced $\sim \,100 \mu $m apart. These cracks proved to have no significant effect on either the loss characteristics or the spectral location of the SPRs. \\
Before filling with metal, the PCF structure is itself a waveguide with the property of being ``endlessly single-mode'' when $ d / \Lambda  < 0.4 $, i.e., under this condition it supports only the fundamental guided mode at all wavelengths where the glass remains transparent \cite{Birks_1997}. The modal refractive index is $< 1$, with the result that the fields are evanescent in the hollow regions, which act as barriers to the propagation of light. The mode may be viewed as being trapped because it has a single transverse lobe that is too large to ``squeeze between'' these barriers, unlike the higher-order modes, which are able to leak away \cite{Russell_2006}. This behavior also occurs in situations where the air channels are replaced with rods of high index glass, or with more complex microstructured features \cite{Stone_2006}. Within certain wavelength ranges these cladding rods can become resonant with the core mode and the light leaks rapidly away. \\
This picture is useful for understanding what happens when light is trapped within a nanowire array (Fig. \ref{fig2}). In the absence of any SPRs, the nanowires strongly reject light, which is unable to propagate through the bulk metal, and light is guided in the core. At certain frequencies, however, SPRs phase-match to the guided mode and light leaks away. \\
To understand the nature of these SPRs in simple terms, we assume that the fields of the surface plasmon-polaritons on opposite sides of an individual nanowire do not interact, but simply spiral around the surface of the wires (Fig. \ref{fig2}). This is valid when the wires are much thicker than the skin depth in the metal. At fixed optical frequency, only a certain discrete number of SPRs exist on the nanowires, each with an integral number of field nodes around the circumference of the wires. Taking into account the dispersion relation of a planar surface plasmon-polariton, it is straightforward to show that the SPRs on the wires have an axial refractive index that approximately follows the relationship:

\begin{equation}
	n_m = \sqrt{ \frac{\epsilon \, \epsilon_M}{\epsilon + \epsilon _M} - \left( \frac{(m-1)\lambda}{d \, \pi} \right)^2}
	\label{equ1}
\end{equation}

where m is the SPR mode order, $\lambda$ the vacuum wavelength, $\epsilon(\lambda)$ and $\epsilon_M(\lambda)$ are the dielectric constants of glass and metal respectively, and $d$ is the diameter of the nanowires. The mode order in Eq. (\ref{equ1}) is reduced by 1 in order to take account of the geometrical phase change around the perimeter of the wires ($\pi$ per $180^\circ$). This geometrical phase makes it impossible to build up a fundamental mode (which must have a constant transverse phase) unless a non-zero azimuthal wavevector is added to cancel it.

\begin{figure}[htb]
	\includegraphics*[width=7.1cm]{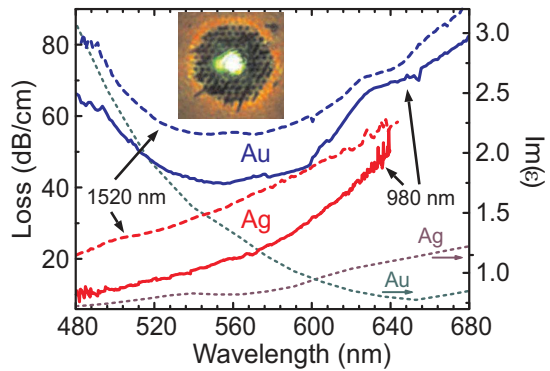}
	\caption{\label{fig3}Loss spectra in the visible for two different PCFs (hole diameters 980 nm and 1520 nm) filled with gold and silver. Also plotted is the imaginary part of the dielectric constant for gold and silver \cite{Palik_1998}. The inset is a near-field optical micrograph of the end-face of a gold-filled PCF when white light is launched into it; the core light is green and the nanowire array is dark, with a bright orange ring surrounding it. The distortions in the picture are caused by irregularities in the cleaved surface.}
\end{figure}

\begin{figure}[htb]
	\includegraphics*[width=7cm]{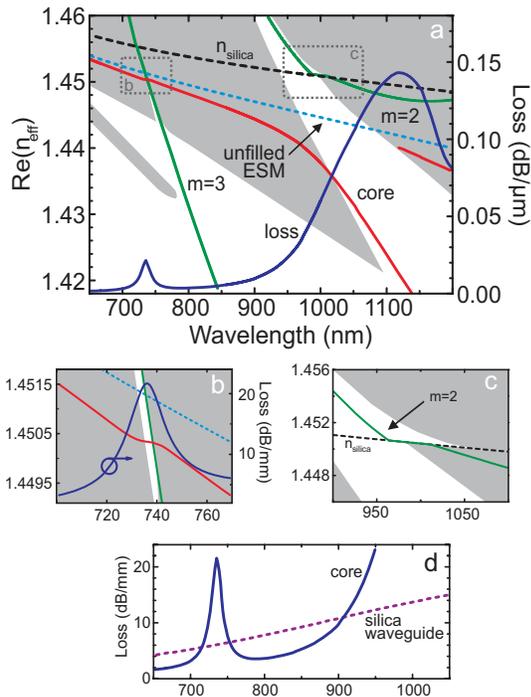}
	\caption{\label{fig4}Dispersion and loss characteristics of the main modes in the system. (a) Wavelength dependence of effective axial refractive index and loss (dark blue curve) for a silver-filled PCF ($d = 1000 nm $, $ \Lambda  = 2.9 \mu $m, with three rings of holes). The green curves are the $m = 2$ and $m = 3$ SPRs for a single wire; the light blue dashed curve represents the guided mode in an unfilled PCF; the dashed black curve is for silica glass; and the red curves are for the guided mode in the nanowire array. The array has photonic band gaps in the grey shaded regions. (b) Close-up view of the $m = 3$ anti-crossing point. (c) Dispersion of the $m = 2$ mode in the regime of the nanowire cut-off. d, Loss characteristics of the fundamental core-mode of the silver-filled PCF compared with the loss of a silver-clad cylindrical silica rod-waveguide of the same diameter ($5.3 \mu $m).}
\end{figure}

When the real part of $n_m$ coincides with the index of the guided mode, energy will resonantly tunnel from the core mode into the nanowires, resulting in strong leakage of power and loss of guidance. When the mode order is high enough, i.e., the angle of spiral is steep enough, $n_m$ becomes less than the index of the silica host and the SPP mode becomes leaky. Comparisons with recently reported numerical results using the multipole method \cite{Poulton_2007} show that the wavelengths at which this occurs are predicted by Eq. (\ref{equ1}) to better than $3\%$ accuracy for wire diameters greater than 500 nm (note that model is only accurate for $m > 0$). The general conclusion is that metallic nanowires with diameters $\sim \,1 \mu $m support spectrally separated SPRs in the visible regime, as predicted by Eq. (\ref{equ1}). \\
Transmission spectra for two identical solid-core PCFs (hole diameter 980 nm), filled with Au and Ag, are presented in the inset in Fig. \ref{fig2}. Clearly visible are dips in transmission at 940 nm for Au and at 920 nm for Ag, caused by phase-matched coupling of the core light to SPRs on the nanowires. By filling and characterizing two further PCFs (the points labeled 1 to 3 in Fig. \ref{fig2}), we established that the dip wavelength scales roughly linearly with nanowire thickness. For the largest hole diameter (1520 nm) the losses were extremely high if all the holes were filled, due to the high light-metal overlap, so to reduce the loss to measurable proportions only three holes were filled in this case. \\
The transmission loss in the visible, which is very high, was measured by successive cut-backs of the metal-filled PCF section. For example, three careful cleaves (to 0.53, 0.36 and 0.24 cm) of an initially short (0.70 cm) section of a gold-filled PCF (d = 980 nm) were made, the transmitted spectrum being measured at each stage. This approach provided three independent estimates of the loss, allowing us to verify the accuracy of the measurement. The resulting spectra for two different gold and silver arrays are plotted in Fig. \ref{fig3}.\\
An optical micrograph of the near-field pattern at the gold-filled end of a PCF, excited from the far end with white supercontinuum light, is shown in the inset in Fig. \ref{fig3}. The core light has a distinct green color, as expected from the position of the transmission band in Fig. \ref{fig3}. Just outside the nanowire array is a ring of orange-colored light, caused by small-angle leakage through SPRs in the nanowires.\\
A convenient starting point for understanding the observed optical behavior is a plot comparing the dispersion characteristics of all the individual modes and materials in the system. Figures \ref{fig4}(a) to (c) show the dependence of effective index on wavelength for the SPRs on a single silver nanowire (the green curves), the guided mode in the glass core (the red and blue curves represent refractive index and loss), the guided mode in an unfilled PCF (the dashed light blue curves), the refractive index of silica glass (the black dashed curve) and the positions of the photonic band gaps in the nanowire array (the grey shaded regions). The calculations were carried out using a multipole expansion of the fields \cite{Poulton_2007,White_2002}. In the nanowire array, pass-bands form around the dispersion curve for a single nanowire, their width being proportional to the inter-wire coupling strength. Thus, as the SPRs approach and cross the silica line from higher effective indices (at shorter wavelengths), their fields spread out, the nanowires couple more strongly, and the pass-bands widen. \\
The silica line marks the boundary between bound and leaky SPP modes on the wires. Since the core-guided mode has an effective index that is always less than that of silica, any SPR that is phase-matched to it will be leaky (the wavevector of the light in silica will be real-valued); it will have the characteristics of a Mie resonance. Phase-matching between the core mode and the SPRs occurs at specific anti-crossing points, in the vicinity of which the optical attenuation is strongly enhanced because power is able to leak away laterally through the wires and escape into the outer glass cladding. 

The experimental results show an overall decrease in transmission at longer wavelengths (Fig. \ref{fig2}, inset), caused by the intrinsic absorption of the metal. Within the photonic band gaps, additional losses arise because of the finite extent of the nanowire array, which permits light to tunnel through to the outer silica cladding via evanescent fields. Such confinement losses can be reduced to negligible proportions by making the nanowire array sufficiently large; for example, in a silver array with $ d / \Lambda  = 0.15 $, numerical modelling shows that 3 rings are sufficient to ensure that the attenuation is dominated by loss in the metal and not by leakage \cite{Poulton_2007}. The green transmission band in the gold-filled PCF (Fig. \ref{fig3}) is explained by metallic absorption on the short wavelength side, and SPRs at longer wavelengths. The wavelength dependence of Re($\epsilon$ ) differs little between gold and silver, resulting in transmission spectra that are quite similar in each case, for identical arrays (Fig. \ref{fig2}, inset). In contrast, Im($\epsilon$) is significantly different for the two metals, the gold arrays exhibiting higher absorption-related losses in the green than the silver (Fig. \ref{fig3}). This is particularly pronounced for $ \lambda < 600 $nm. \\
As the nanowire thickness increases at fixed pitch, the transmission loss rises because of the larger light-metal overlap, and the coupling between SPRs also increases, causing the pass-bands and the transmission dips to widen as one moves from point 1 to 3 in Fig. \ref{fig2}. This was observed in the experimental spectra (not shown here). The experimental points (Fig. \ref{fig2}) lie between the theoretical curves for the $m = 2$ and $m = 3$ SPRs, being closer to the $m = 2$ SPR for smaller nanowires.

This we attribute to air-gaps between the nanowires and the glass, caused by the higher thermal expansion coefficient of the metal. Assuming an azimuthally constant gap, simulations show that the shift to shorter wavelength has an almost linear dependence on its width (e.g., 18 nm/nm for the $m = 2$ dip in gold nanowires 1000 nm in diameter, when the SPR wavelength is $\sim 1000 $ nm). The discrepancy between measured and simulated data improves for narrower wires because the gap size is smaller (a simple analysis predicts gap widths of 4, 7 and 11 nm for gold wires 550, 980 and 1520 nm in diameter, and $\sim 20\%$ larger for silver), as is apparent in Fig. \ref{fig2}.\\
In Fig. \ref{fig3}, the minimum loss wavelength in the gold-filled PCFs is shifted $\sim 100$ nm to shorter wavelengths compared to bulk metal. This is because the increasing value of Im($\epsilon$ ) at shorter wavelengths is counterbalanced by a decreasing light-metal overlap integral. Both gold loss curves exhibit a kink in shape at around 640 nm. This effect correlates to the spectral dependence of the skin depth, which shows a distinct local maximum at 640 nm.\\
Away from the SPRs, the transmission loss in a nanowire array is significantly lower than that of a silica rod of comparable size embedded in silver (Fig. \ref{fig4}(d)). For instance at $\lambda$ = 800 nm, a 2 mm long device has a transmission loss of 5 dB, compared to 15 dB for a silica rod in silver. With pitch and wire diameter 10 $\mu$m and 1.5 $\mu$m respectively, the guidance loss in a silver array can be reduced to 0.17 dB/mm at $\lambda$ = 1.55 $\mu$m \cite{Poulton_2007}. \\
In conclusion, high quality nanowires can be produced by pumping molten metal into the hollow channels of silica glass PCF. Within the constraints of the pressurized filling technique, nanowire diameters of 200 nm appear possible, which will shift the SPRs to wavelengths in the range of lowest overall absorption. Such nanowires have many potential applications, including in electrochemistry, glass poling, liquid crystal devices and imaging on the sub-wavelength scale.

\end{document}